\documentclass[a4paper]{article}

\usepackage{icrc2013}

\title{High-energy neutrino production from photo-hadronic interactions of gamma rays from 
Active Galactic Nuclei at source}

\shorttitle{HE $\nu$ production from photo-hadronic interactions of $\gamma$ rays from AGN's 
at source}
 
\authors{
J.C.~Arteaga-Vel\'azquez$^{1}$, 
Angelo Mart\'\i nez $^{2}$ 
}

\afiliations{
$^1$ Instituto de F\'{\i}sica y Matem\'aticas, Universidad Michoacana, Morelia, Michoacan, 
Mexico\\
$^2$ Facultad de Ciencias  F\'{\i}sico-Matem\'aticas, Universidad Michoacana, Morelia, Michoacan,
Mexico\\
}

\email{arteaga@ifm.umich.mx}

\abstract{Recent astronomical observations reveal that Active Galactic Nuclei (AGN) are sources
 of high-energy radiation. For example, the \textit{Fermi-LAT} and \textit{Hess} telescopes 
 have detected gamma-ray emissions from the cores of several types of AGN's. Even more, the 
 Pierre Auger observatory has found a correlation of ultra-high energy cosmic ray events with 
 the position of Active Galactic Nuclei, such as Centaurus A. In this way, according to particle 
 physics, a flux of high-energy neutrinos should be expected from the interactions of cosmic and 
 gamma-rays with the ambient matter and radiation at the source. In this work, estimations of 
 the diffuse neutrino flux from AGN's arising from interactions of the gamma radiation with the 
 gas and dust of the sources will be presented.}


\begin{document}
\maketitle

\section{Introduction}

 Active Galactic Nuclei are among the most luminous extragalactic objects in the universe.
 Their emission covers all the electromagnetic spectrum, from radio up to the gamma-ray
 wavelengths and seems to have its origin in gravitational potential energy of matter 
 falling inwards a supermassive black hole hidden deep at the source. It is around these
 supermassive objects where suitable conditions for particle acceleration may be found. 
 Nowadays, it is believed that this process is behind the
 origin of the $\gamma$-ray emission seen from AGN's. In fact, there are two known particle
 physics scenarios for the production of $\gamma$-rays in AGN's, the so called hadronic
 and leptonic models. The former  involves the acceleration and production of
 protons and atomic nuclei, whose interactions with the matter and radiation at the
 source lead to the production of secondary hadrons, which decay in $\gamma$-photons and 
 neutrinos  \cite{Becker}. In this case, active galactic nuclei could also be important
 sources of high-energy neutrinos and cosmic rays. On the other hand, in the framework
 of the leptonic model, electrons and positrons are accelerated up to the relativistic
 regime producing gamma rays by radiative processes \cite{Bottcher}. Although, no hadron  
 acceleration up to the highest energies occurs, neutrinos are still produced in this 
 scenario by means of secondary mechanisms induced by $\gamma$-ray collisions with the 
 material and radiation of the source. In any case,  no matter the mechanism of 
 $\gamma$-ray production, the sole presence of this radiation implies the existence of 
 at least a feeble flux of high-energy neutrinos coming from gamma-ray interactions at 
 the source and its surroundings.

  In reference \cite{Arteaga}, the diffuse flux of neutrinos from FR I and BL Lac type 
 galaxies produced by photo-hadronic interactions of gamma rays during their way out 
 the source and the host galaxy was investigated. In that work, in fact, it was shown that 
 this particular  flux is out of the reach of modern neutrino telescopes. Even more, in case 
 that bigger neutrino detectors could be built, the detection of this diffuse $\nu$ flux 
 would be difficult to achieve due to the presence of the strong atmospheric $\nu$ background. 
 However, it was learned that high-energy neutrino emission is not absent in AGN's at all 
 and that, if neutrinos are ever detected from FRI and BL Lac galaxies, they would come 
 from hadronic scenarios. In the present contribution, the research is extended to 
 FR II galaxies. These objects have a bigger intrinsic absorption in X-rays than 
 FR I galaxies \cite{Evans}, therefore, they offer more target material 
 for $\nu$-production by photo-hadronic interactions. Besides, from the \textit{Fermi-LAT} 
 surveys, it results that FR II galaxies with $\mbox{MeV} - \mbox{TeV}$ emission are 
 more $\gamma$-ray luminous than FRI type AGN's, which can also enhance  
 neutrino emission by $\gamma$-ray collisions in those environments. However, their
 contribution to the diffuse flux of high-energy neutrinos in the universe could be 
 strongly compromised due to the fact that the population of FR II seems to escape 
 from the \textit{Fermi-LAT} detection \cite{FERMI}, which may be caused by an 
 anisotropic emission from the source. If true, this phenomenon would favor the 
 detection of FR II type galaxies with small jet inclination angles with respect to 
 the observer's line of sight in the \textit{Fermi-LAT} data \cite{FERMI}. The paper 
 is organized as follows: First, a brief description of the calculation of the 
 diffuse flux of neutrinos from individual sources is presented. Then, the $\gamma$-ray
 spectral luminosity of FR II objects is shown along with the associated neutrino
 flux for a single source. Next, a model for the matter distribution of the powerful
 radio galaxy Cygnus A is described. This is the closest FR II object to the Earth.
 And finally, the diffuse flux of neutrinos from photo-hadronic interactions of 
 $\gamma$-photons from FR II galaxies is given.\vspace{-0.5pc}  

 \section{The diffuse neutrino flux}

 The extragalactic flux of neutrinos (in units of $\mbox{s}^{-1} \cdot \mbox{sr}^{-1}
 \cdot \mbox{TeV}^{-1} \cdot \mbox{cm}^{-2}$) detected at Earth is estimated from the 
 following expression:
 \begin{eqnarray}
    \frac{d\Phi_\nu(E^{\circ}_\nu)}{d\Omega^{\circ}} 
      &=&
      \frac{c}{4\pi} \int_{0}^{z_{max}} \frac{dz}{H(z)}
      \int_{\log_{10}\mathcal{L}_{\gamma}^{min}}^{\log_{10}\mathcal{L}_{\gamma}^{max}}
       d(\log_{10}\mathcal{L}_\gamma) \nonumber \\ 
      &&
      \cdot \rho_\gamma(\mathcal{L}_\gamma, z)\cdot
      L_\nu[\mathcal{L}_\gamma, E^{\circ}_\nu(1+z)], 
    \label{eq1}
 \end{eqnarray}
 where $ L_\nu[\mathcal{L}_\gamma, E_\nu]$ is the neutrino spectral luminosity of 
 a FR II type source localized at redshift $z$ and characterized by an integrated 
 $\gamma$-ray luminosity $\mathcal{L}_\gamma$ in the interval from $100 \, \mbox{MeV}$
 to $10 \, \mbox{GeV}$. Here, $E_\nu =  E^{\circ}_\nu(1+z)$ represents the neutrino energy 
 at source and $E^{\circ}_\nu$, the redshifted energy as measured at Earth. On the other 
 hand, $\rho_\gamma(\mathcal{L}_\gamma, z)$ is the gamma-ray luminosity function (GLF) 
 of FR II sources per comoving volume $dV_c$ and interval $d(\log_{10}\mathcal{L}_\gamma)$, 
 as given by \cite{Inoue}. The Hubble parameter at $z$ is represented by $H(z)$. Along 
 the paper, a $\Lambda$CDM cosmology is assumed with $\Omega_\Lambda = 1 - \Omega_m =
 0.74$. Integral limits are $z_{max} = 5$  \cite{Inoue} and 
 $\mathcal{L}_{\gamma}^{min (max)} = 43 (50) \, \mbox{ergs}/\mbox{s}^{-1}$ (based on
 observations from FR II and FSRQ type objects performed by \textit{Fermi}-LAT \cite{FERMI}.
 \vspace{-1.5pc}

 \begin{table}[!t]
\footnotesize
\begin{center}
\begin{tabular}{l|c}
\hline
\hline
   AGN & $\alpha$ \\
\hline
  3C380 & $-2.51 \pm 0.30$\\
  3C207 & $-2.42 \pm 0.10$\\
  3C111 & $-2.54 \pm 0.19$\\
  PKS0943 & $-2.83 \pm 0.16$\\
\hline
\hline
\end{tabular}
\caption{Spectral indexes for the gamma-ray fluxes measured by \textit{Fermi-LAT} from
several FR II type objects \cite{FERMI}.}
\label{tab1}
\end{center}
\vspace{-1.5pc}
\end{table}

 \section{Luminosities of FR II galaxies}

 As in reference \cite{FERMI}, we assumed a power-law spectrum for the photon spectral
 luminosity at source, $L_\gamma(E_\gamma) = dN_\gamma/dt dE_\gamma = 
 L \cdot E_\gamma^{\alpha}$. Measurements of the spectral slope, $\alpha$, in the
 interval $100 \, \mbox{MeV} - 10 \, \mbox{GeV}$ were provided in 
 \cite{FERMI} by the \textit{Fermi-LAT} collaboration for a set of four FR II galaxies 
 (see table \ref{tab1}). For our calculations, we adopted a standard FR II source with 
 $\alpha = -2.57 \pm 0.18$, which corresponds to the mean value of the spectral index of 
 the set of data in table \ref{tab1} (the error represents the corresponding standard 
 deviation of the data). We also add an energy cutoff around $E = 100 \, \mbox{TeV}$, 
 assuming conservatively that the leptonic model is at work. On the other hand, we 
 will adopt the hypothesis that the $\gamma$-ray emission from FR II type galaxies 
 is anisotropic to explain the small number of FR II objects detected by \textit{Fermi-LAT} 
 \cite{FERMI}. The main idea behind this hypothesis is that the $\gamma$-ray flux is 
 born in the AGN jet as a result of Compton scattering of external photons by electrons, 
 which produces a strong Doppler boosting and a narrow beaming cone of emission \cite{FERMI}. 
 From the discussion at the beginning of this section and following \cite{Dermer, Dermer2}, 
 the photon spectral luminosity per solid angle interval at the reference frame of the 
 source galaxy will have the following form:
  \begin{equation}
   \frac{dL_\gamma(E_\gamma, \theta, i)}{d\Omega} = N \cdot \eta(\theta, i) 
   \cdot \left( \frac{E_\gamma}{\mbox{TeV}} \right)^{\alpha} \cdot 
   e^{-(E_\gamma/10^{2} \, \mbox{\footnotesize{TeV}})},
   \label{eq2}
 \end{equation}
 where $i$ is the angle between the jet direction of the AGN and the line of sight to the 
 observer and $\theta$, the angle between the jet axis and the direction of the emitted 
 photon. Here, $N$ is a normalization factor chosen in such a way that for an observer
 with an angle of view $i$, the measured integrated luminosity per interval
 of solid angle at source that would be measured in that direction is 
 $\mathcal{L}_\gamma/4\pi$. On the other hand,
 \begin{equation}
  \eta(\theta, i)  =\frac{1}{4\pi} \left[\frac{\delta_D(\theta)}{\delta_{D}(i)}  
   \right]^{4 + 2a} \left( \frac{\mu_i}{\mu_\theta}  
   \right)  \left( \frac{1 + \mu_\theta}{1 + \mu_i} \right)^{2 + a},
  \label{eqn3}
 \end{equation}
  with $\delta_D(\theta) = 1/[\Gamma (1 - \mu_{\theta} \beta)]$, the Doppler factor, and
  $\mu_{\theta} = \cos (\theta)$.  Similar expressions apply for $\mu_i$ and  
 $\delta_D(i)$. In the above equation, $\Gamma$ is the Lorentz factor of the material
 in the jet and $\beta = \sqrt{1 - 1/\Gamma^2}$. We will take $\Gamma = 5$ and $10$.
 Finally, $a = -\alpha - 1$.

 For the calculation of the neutrino luminosity, only photo-hadronic interactions
 are taken into account. Contributions from  $\mu$-pair production in $\gamma$-ray
 interactions with matter and photons will be neglected due to their 
 lowest cross-section. The target for the gamma radiation will be the nucleons of the
 gas and dust of the AGN and its host galaxy. We will assume that this material is 
 composed of protons with energies well below $100 \, \mbox{MeV}$. In this way, they will
 be considered at rest during the calculations of the $\gamma P$ collisions. The neutrino 
 spectral luminosity per solid angle interval along a given direction $\theta$ from the jet 
 axis is given by the expression
 \begin{eqnarray}
     \frac{dL_\nu(E_\nu,  \theta, i)}{d\Omega} dE_\nu &=&
     \Sigma_{\small H}(\theta) \int_{E_{\gamma, i}}^{E_{\gamma, f}} dE_\gamma
    Y^{\gamma P \rightarrow \nu}(E_\gamma, E_\nu) \nonumber \\
    && \cdot \sigma_{\gamma P}(E_\gamma)  dL_\gamma (E_\gamma, \theta, i)/d\Omega,
    \label{eq4}
  \end{eqnarray}
 when the observer has an angle of view, $i$, with respect to the jet direction.
 Here, $\Sigma_{\small H}(\theta)$ is the column density of target protons in the
 direction $\theta$, $ \sigma_{\gamma P}(E_\gamma)$ is the $\gamma P$ cross-section at
 a photon energy $E_\gamma$ and $Y^{\gamma P \rightarrow \nu}(E_\gamma, E_\nu)$ is the
 $\nu$ yield, i.e., the number of neutrinos produced with energy around $dE_\nu$
 during a collision of a $\gamma$-ray with energy in the interval $dE_\gamma$ with
 a proton at rest. The cross-section $ \sigma_{\gamma P}(E_\gamma)$ was evaluated
 according to \cite{PDG} and the yield of neutrinos was taken from \cite{Arteaga},
 where it was calculated with the Monte Carlo program SOPHIA v2.01 \cite{Sophia}.

  Using equation \ref{eq2} in expression \ref{eq4}, summing over all directions 
 $\theta$ and averaging on the observer's view angle, $i$, we arrive to the formula
 \begin{eqnarray}
      L_\nu(E_\nu)dE_\nu &=& N \cdot \xi(\mathcal{L}_\gamma)  
      \int_{E_{\gamma, i}}^{E_{\gamma, f}} dE_\gamma
      Y^{\gamma P \rightarrow \nu}(E_\gamma, E_\nu) \nonumber \\
     && \cdot \sigma_{\gamma P}(E_\gamma) \left( \frac{E_\gamma}{\mbox{TeV}} \right)^{\alpha} 
   e^{-(E_\gamma/10^{2} \, \mbox{\footnotesize{TeV}})},
    \label{eq5}
 \end{eqnarray}
 where
 \begin{equation}
   \xi(\mathcal{L}_\gamma)  = \frac{
    \int_{\delta \Omega(\mathcal{L}_\gamma)}
    \int_{4\pi \footnotesize{sr}} 
    d\Omega_\theta d\Omega_i \Sigma_{\small H}(\theta) \cdot \eta(\theta, i)}
    {\delta \Omega(\mathcal{L}_\gamma)}.
   \label{eq6}
 \end{equation}
 In the above formula, we have put a constraint to the direction $i$ of the observer.
 This restriction is obtained when the luminosity of FR II galaxies, observed 
 from the direction $i$ with photon spectral luminosity $\mathcal{L}_\gamma$, is
 limited to be smaller than   $10^{50} \, \mbox{ergs}/\mbox{s}$ along the jet axis. 
 This upper limit comes from the observations of FSRQ galaxies with the \textit{Fermi-LAT} 
 telescope \cite{FERMI}, which are just FR II AGN's observed along the jet direction 
 according to the unified AGN model \cite{Urry}. In this way, we integrate the angle 
 $i$ only inside a limited solid angle interval $\delta \Omega(\mathcal{L}_\gamma)$ 
 for which the aforementioned condition is valid. Finally, in equation \ref{eq5}, 
 integration is performed from $E_\gamma = 10^{-0.8}$ (just above the $\gamma$-energy 
 threshold for pion photo-production) to $10^{6} \, \mbox{GeV}$. \vspace{-1.5pc}

 \begin{table*}[!t]
\footnotesize
\begin{center}
\begin{tabular}{l|c|c}
\hline
\hline
   Structure & $R[pc]$ & $n_H [\mbox{cm}^{-3}]$ \\
\hline
  Core & $0.03$ & $10^{6}$\\
  BLR region & $0.03 - 0.6$ & $ (7.3 \times 10^{4}) \cdot  e^{- (\beta/20^{\circ})^2}$ \\
  Torus & $0.6 - 130$ & $ (3.2 \times 10^{2}) \cdot e^{- (\beta/20^{\circ})^2}$ \\
  Inclined disk & $0.03 - 80$ & $10^{4}$\\
  Dust lane & $(0.13 - 1.5)\times 10^{3}$ & $5.6 \times 10^{-1}$\\
  ISM & $2 \times 10^{3}$ & $0.11/[1 + (7.69 \times 10^{-7}) \cdot r^2]^{0.946}$ \\
  Cocoon & $r_{-} = 15  \times 10^{3}$; $r_{+} = 60  \times 10^{3}$   & $1.9 \times 10^{-4}$ \\
  Shell  & $r_{-} = 25  \times 10^{3}$; $r_{+} = 62.5\times 10^{3}$ & $1.4 \times 10^{-2}$ \\
  Halo   & $5 \times 10^5$ & $7.9/r^2$\\
\hline
\hline
\end{tabular}
\caption{Extension and density distributions of the main gas structures in the model 
of Cygnus A. Here, $r$ represents the spherical radius, $r_{-}$, the semi-minor axis, and
$r_{+}$, the semi-major axis of the prolate spheroidal structures. Meanwhile, $\beta$
 is the angle measured from the equatorial plane of the galaxy. It can have only
 values in the interval $[-20^{\circ}, 20^{\circ}]$.}
\label{tab2}
\end{center}
\vspace{-1.5pc}
\end{table*}

 \section{Column density}

 To estimate the column density of the gas and dust at the source, first, we assume
 that the birth place of the gamma radiation observed from FR II type galaxies
 is found at the nucleus of the AGN and coincides with the radio core location. 
 In fact, combined multi-wavelength observations of the radio galaxy M87, which hosts 
 a FR I type object, point out that the gamma-ray production site could be located 
 at the nucleus of the AGN \cite{m87-multi09}. On the other hand, we take Cygnus A as a model
 for FR II galaxies. The advantage of this choice is that this  powerful FR II galaxy
 has been well studied by different astronomical instruments due to its proximity to 
 the Earth ($z = 0.056$) \cite{Smith}. Based on reported astronomical observations,
 we built a simple model of Cygnus A to describe the matter distribution of the
 main structures of the galaxy (see table \ref{tab2}). 

 To start with, for the radius of the gamma-source at the nucleus, we take the limit
 derived from \cite{Barvainis} for the compact radio source in Cygnus A. The density 
 was assumed to be $10^{6} \, \mbox{cm}^{-3}$, which agrees with estimations at the 
 nucleus of AGN's \cite{NucleusDensity}. Observations against the nucleus indicates 
 the presence of a strong X-ray absorber, which could be due to material from 
 the BLR region and a dusty torus. In \cite{Privon}, the torus was modeled with a
 clumpy circumnnuclear disk. Its geometry and density were restricted with 
 radio and IR data. For the torus, we will use the disk model of \cite{Privon} with a 
 constant radial density, only depending on the angle $\beta$ respect to the equator 
 of the galaxy, a  half opening angle $\sigma = 20^{\circ}$ (measured from the 
 equatorial plane), an internal radius equal to $0.6 \, \mbox{pc}$, an outer 
 radius of $130 \, \mbox{pc}$ and with axis oriented along the jet direction. We will 
 assume that the BLR region is an inner extension of the torus \cite{Nemov}. Therefore,
 the BLR region will be described also with a disk with the same geometrical parameters, but with 
 smaller dimensions, particularly with an external radius of $0.6 \, \mbox{pc}$. Density is 
 found in such a way that for $\beta = 10^{\circ}$,  which corresponds to the viewing angle 
 of an observer at the Earth \cite{Privon}, the total column depth along the BLR and torus 
 region is $N_H = 2 \times 10^{23} \mbox{cm}^{-2}$, the value reported by the \textit{Chandra} 
 telescope for the X-ray absorber \cite{Young}. We will assume that this column density is 
 equally divided between the BLR region and the torus, since it is unknown exactly what is 
 the exact contribution from each region to the X-ray absorption. 
 
 A second disk, tilted $21^{\circ}$ with respect to the equatorial plane of the host galaxy, 
 a density $n_H = 10^{4} \mbox{cm}^{-3}$, a radius of $80 \, \mbox{pc}$ and an opening
 angle of $14^{\circ}$ was also added to our model \cite{Struve}. It was detected with
 VLBA HI absorption studies of the core region of Cygnus A \cite{Struve}. 

 Observations at IR wavelengths also reveal an edge-on oriented biconical structure, which is 
 likely caused by a kpc-scale dust lane characterized by a disk geometry and funnels along 
 the jet axis \cite{Tadhunter99}. In the model of \cite{Tadhunter99}, the funnels have an 
 opening angle of $116^{\circ}$, while the disk axis is aligned within $15^{\circ}$ with the 
 jet direction. Here, we will incorporate the dust lane with the above configuration in 
 our model. Besides, for simplicity, we will consider that the disk lies on the equatorial 
 plane of the galaxy. The external diameter and the width of the disk are set to $3 \, 
 \mbox{kpc}$ \cite{Bellamy} and $1.5 \, \mbox{kpc}$ \cite{Young}. On the other hand, we will 
 assume that the inner frontier of the disk coincides with the external radius of the torus.
 To calculate the density, we take the column density estimated in \cite{Young}
 for the dust lane along the direction to Earth. Taking into account the geometry of
 the disk and the orientation of the observer, and assuming an homogeneous distribution
 of the material inside the dust lane, we estimate a density of $5.6 \times
 10^{-1} \, \mbox{cm}^{-3}$.
  
 The distribution of the interstellar matter up to $2 \, \mbox{kpc}$  will be described 
 using the same function derived in \cite{Tadhunter}, from optical and IR observations, 
 for the stellar density profile in Cygnus A within the $2 \, \mbox{kpc}$ region around
 its nucleus. As in \cite{Tadhunter}, we will assume that matter is spherically distributed 
 in this zone. 

 Observations suggest that the above structures are surrounded by a cocoon and an external shell
 with shocked matter \cite{Carilli}. The shapes and densities of these structures were modeled 
 in \cite{Mathews}. Here, we will take the model number 3 of \cite{Mathews}, which seems to 
 be in better agreement with observations. According to \cite{Mathews}, these structures 
 have a prolate spheroidal shape. The inner (outer) radius of the shell is of the order of 
 $60 (62.5)  \, \mbox{kpc}$ along the semi-major axis, according to the above model, while
 along the semi-minor axis is $15 (25) \, \mbox{kpc}$. On the other hand, in model 3 of 
 \cite{Mathews}, the cocoon has a volume of $2.8 \times 10^4 \, \mbox{kpc}^3$ and encloses 
 a mass of $1.4 \times 10^{8} \, M_\odot$. From this data the mean cocoon's density is 
 estimated. It is worth to mention that the density of the  interstellar region at $2 \, 
 \mbox{kpc}$ is normalized in such a way that it reproduces the density of the cocoon. 
 An average density of $1.4 \times 10^{-2} \mbox{cm}^{-3}$ for the shocked region was estimated 
 from the 2D density plots of \cite{Mathews}.
  
 Beyond the cocoon and the shell, we found the galactic halo, which is composed of hot and
 low density gas. This structure has been detected in X-rays. Its emission extends up to 
 $720  \, \mbox{kpc}$ according to \cite{Smith}. The electron density has been modeled in 
 \cite{Smith} from $60\, \mbox{kpc}$ to $500  \, \mbox{kpc}$ from X-ray observations of the 
 halo of Cygnus A. We will use this profile for our proton density, assuming that the
 proton and electron densities are similar inside the halo. We will apply such an expression
 only up to $500  \, \mbox{kpc}$. \vspace{-0.5pc}

 \begin{figure}[!t]
  \centering
  \includegraphics[height=2.0in, width=3.2in]{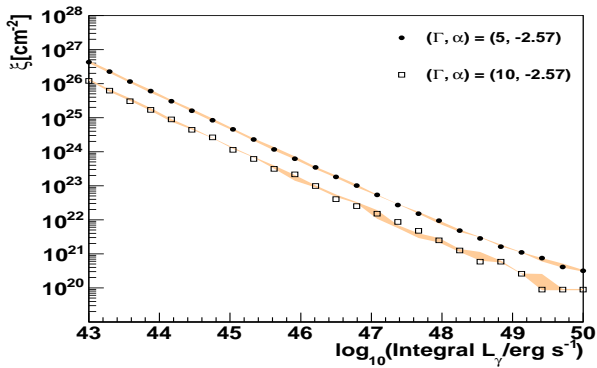}
  \caption{$\xi$ factor for $\alpha = -2.57$ and $\Gamma = 5$ (circles) and
  $10$ (squares). The bands are generated when the photon index is varied inside
  its corresponding error interval.}
  \label{Fig01}
 \vspace{-1pc}
 \end{figure}

 \section{Results and Discussion}

\begin{figure}[!t]
  \centering
  \includegraphics[height=2.0in, width=3.2in]{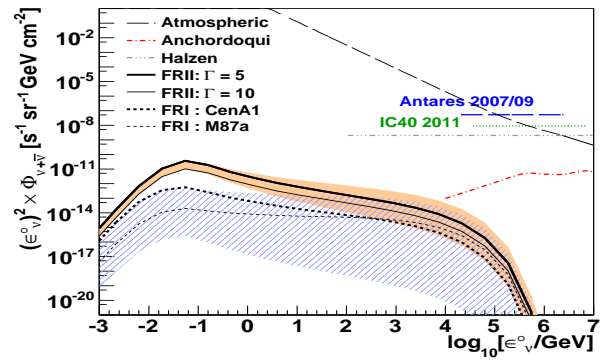}
  \caption{Diffuse flux of neutrinos (for each type) expected from FRII galaxies 
           assuming different Lorentz factors ($\Gamma = 5, 10$) for the AGN jets.The 
           shadowed band covers the results obtained by varying the  photon index inside
           its error interval for the considered $\Gamma$ factors. The flux is compared 
           with the expected results for FRI type objects based on a Centaurus A and a 
           M87 models (dotted black lines and hatched area) \cite{Arteaga}. $\nu$ oscillation 
           is taken into account. The $90 \%$ C.L. upper limits on the diffuse $\nu$ flux 
           derived by the \textit{ICECUBE} \cite{ic402011b} and the \textit{Antares}
           \cite{antares2011b} collaborations are shown. For comparison, the
           diffuse flux of atmospheric neutrinos (segmented line)\cite{Gandhi98},
           including the prompt component from $D$-meson decays \cite{Volkova80},
           is also presented. Two predictions for the diffuse $\nu$ flux from AGN's
           using hadronic  models (Anchordoqui \cite{Anchordoqui08} and Halzen
           \cite{halzen08}) are included in the plot.}
  \label{Fig02}
 \vspace{-1pc}
 \end{figure}

 The $\xi(\mathcal{L}_\gamma)$ factor is presented in figure \ref{Fig01} for different 
 $\Gamma$ and $\alpha$ parameters. We notice that the $\xi(\mathcal{L}_\gamma)$ factor 
 decreases as the solid angle $\delta \Omega(\mathcal{L}_\gamma)$ gets smaller for high 
 $\Gamma$ values and $\mathcal{L}_\gamma$ luminosities. As we will see, that will imply 
 low diffuse neutrino fluxes from FR II galaxies for high Lorentz factors. The  
 neutrino background flux as observed at Earth from FR II objects, after $\nu$ 
 oscillations, is shown in figure \ref{Fig02}. The error band of this flux, calculated
 by varying the photon index ($\alpha$) of formula \ref{eq2} in the interval $-2.57 
 \pm 0.18$ is also presented. The spectrum is compared with the background of  atmospheric 
 neutrinos and the result for FR I type sources \cite {Arteaga}. Two fluxes derived 
 from hadronic models \cite{Anchordoqui08, halzen08} and two experimental upper 
 bounds from \textit{ICECUBE}  \cite{ic402011b} and \textit{Antares} \cite{antares2011b},
 respectively, are also shown. The results show that, although $\gamma$-ray FR II emitters 
 seems to be less abundant than the corresponding FR I population, the higher Doppler and beaming
 factors of FR II galaxies can make their corresponding $\nu$ background flux higher than 
 the one for FR I objects. However this difference is not big. Fig. \ref{Fig02} shows that,
 the investigated neutrino flux from FR II galaxies is too small in comparison  with
 the experimental bounds and the atmospheric background of neutrinos. Neither this feeble 
 flux nor the flux from FR I objects can explain the observed neutrino events detected 
 recently by \textit{ICECUBE} around $1 \, \mbox{PeV}$ \cite{Aartsen}. Therefore, if these events
 comes from FR I and  FR II objects, their most probable origin would be found at hadronic 
 mechanisms originated by cosmic-ray acceleration. \vspace{-0.5pc}

 \section{Conclusions} 

 We have shown that in addition to FR I AGN's, FR II type galaxies characterized by
 $\gamma$-ray emission also produce a small neutrino diffuse flux produced by 
 photo-hadronic interactions by $\gamma$ radiation with the gas and dust at the sources.
 Both fluxes, in general, are too low to be detected by the modern neutrino telescopes and 
 to explain the last $PeV$ neutrino events detected recently by the ICECUBE observatory.

 \vspace{1pc}
 \footnotesize{{\bf Acknowledgment:}{ This work was partially supported by the 
 \textit{Consejo de la Investigaci\'on Cient\'\i fica} of the \textit{Universidad 
 Michoacana} (Project CIC 2012-2013) and \textit{CONACYT} (Project CB-2009-01 132197).}}

\end{document}